\documentclass[12pt]{article}

\usepackage{setspace}
\usepackage{graphicx}
\usepackage{enumerate}
\usepackage{float}
\usepackage[top=1.5in, bottom=1.5in, left=1in, right=1in]{geometry}
\usepackage{hyperref}
\usepackage{rotating}

\usepackage{color}
\usepackage{listings}
\usepackage{morefloats}

\usepackage{amsmath}
\usepackage{verbatim}
\usepackage{cleveref}
\usepackage{pslatex}
\usepackage[justification=centering]{caption}
\usepackage{subcaption}
\usepackage{placeins}
\usepackage[T1]{fontenc}
\usepackage{booktabs,caption,fixltx2e}
\usepackage[flushleft]{threeparttable}

\graphicspath{{/home/lindsayad/gdrive/Pictures/}}

\definecolor{codegreen}{rgb}{0,0.6,0}
\definecolor{codegray}{rgb}{0.5,0.5,0.5}
\definecolor{codepurple}{rgb}{0.58,0,0.82}
\definecolor{backcolour}{rgb}{0.95,0.95,0.92}

\lstdefinestyle{mystyle}{
    backgroundcolor=\color{backcolour},
    commentstyle=\color{codegreen},
    keywordstyle=\color{magenta},
    numberstyle=\tiny\color{codegray},
    stringstyle=\color{codepurple},
    basicstyle=\footnotesize,
    breakatwhitespace=false,
    breaklines=true,
    captionpos=b,
    keepspaces=true,
    numbers=left,
    numbersep=5pt,
    showspaces=false,
    showstringspaces=false,
    showtabs=false,
    tabsize=2
}

\title{Fully Coupled Simulation of the Plasma Liquid Interface and Interfacial Coefficient Effects}
\author{Alexander Lindsay, David Graves, and Steven Shannon}

\begin{document}
\maketitle

\begin{abstract}

There is a growing interest in the study of coupled plasma-liquid systems because of their applications to biomedicine, biological and chemical disinfection, agriculture, and other areas. Optimizing these applications requires a fundamental understanding of the coupling between phases. Though much progress has been made in this regard, there is still more to be done. One area that requires more research is the transport of electrons across the plasma-liquid interface. Some pioneering works (\cite{rumbach2015solvation,rumbach2015effect}) have begun revealing the near-surface liquid characteristics of electrons. However, there has been little work to determine the near-surface gas phase electron characteristics. Without an understanding of the near-surface gas dynamics, modellers are left to make assumptions about the interfacial conditions. For instance it is commonly assumed that the surface loss or sticking coefficient of gas-phase electrons at the interface is equal to 1. In this work we explore the consequences of this assumption and introduce a couple of ways to think about the electron interfacial condition. In one set of simulations we impose a kinetic condition with varying surface loss coefficient on the gas phase interfacial electrons. In a second set of simulations we introduce a Henry's law like condition at the interface in which the gas-phase electron concentration is assumed to be in thermodynamic equilibrium with the liquid-phase electron concentration. It is shown that for a range of electron Henry coefficients spanning a range of known hydrophilic specie Henry coefficients, the gas phase electron density in the anode can vary by orders of magnitude. Varying reflection of electrons by the interface also has consequences for the electron energy profile; increasing reflection may lead to increasing thermalization of electrons depending on choices about the electron energy boundary condition. This variation in anode electron density and energy as a function of the interface characteristics could also lead to significant variation in near-surface gas chemistries when such reactions are included in the model; this could very well in turn affect the reactive species impinging on the liquid surface. We draw the conclusion that in order to make more confident model predictions about plasma-liquid systems, finer scale simulations and/or new experimental techniques must be used to elucidate the near-surface gas phase electron dynamics.

\end{abstract}

\section{Introduction}

In the low-temperature plasma community there is a burgeoning interest in the study of plasma-liquid systems for both basic and applied research purposes. Applications stemming from the interactions of plasmas and liquids include biomedicine and biological disinfection \cite{Kong2009b,Laroussi2009,Shimizu2014c,VonWoedtke2014a,VonWoedtke2013a,Joubert2013a}, chemical disinfection \cite{Johnson2006,Locke2006,Theron2008}, and agricultural uses. \cite{Park2013b,Lindsay2014} To most effectively utilize plasma-liquid systems requires a fundamental knowledge of their behavior; many researchers are now actively contributing to that knowledge through both experimental \cite{Lukes2014b,Bruggeman2009d,Pavlovich2013g,Traylor2011h,yagi2015two,bruggeman2008dc,rumbach2015solvation} and modelling efforts. \cite{Babaeva2014b,Tian2014,Chen2014a,shirafuji2014numerical} Though much progress has been made, there is still much that is unknown, particularly in the interfacial region where the plasma meets the liquid. For instance, little is really known about how electrons are transported across the interface. Most studies in the literature consider solvation of electrons generated in the aqueous bulk by radiolysis. \cite{kimura1994ultrafast,paik2004electrons} A highly energetic electron is ejected from the solvent molecule and is initially delocalized in the solvent's conduction band. Eventually the electron is localized in a solvent trap and is electronically relaxed. The electron relaxation is accompanied by orientation of the solvent molecules to solvate the rapidly changing charge distribution. \cite{kimura1994ultrafast} While this qualitatively explains the behavior of several eV electrons generated in the liquid bulk, researchers are keen to learn what additional physiochemical effects might be associated with electron transport across an interface. Rumbach et. al. \cite{rumbach2015solvation} used absorption spectroscopy to detect the presence of solvated electrons in the surface region with an estimated penetration depth of 2.5 nm. A molecular dynamics study indicates that electrons at the surface of water only have about 10\% of their density protruding into the vapor phase, suggesting that their behavior should be much more characteristic of a fully hydrated as opposed to a half-hydrated species. \cite{uhlig2013electron} These studies help elucidate the character of the liquid phase side of the interface. However, little work has been done to understand the electron behavior on the gas phase side of the interface. Common gas discharge modelling parameters like the surface loss coefficient for electrons are unknown for the gas-liquid interface. To date plasma-liquid models have assumed a surface loss coefficient of unity \cite{Tian2014,shirafuji2014numerical}, however, there is no known molecular scale simulations or experimental measurements to indicate that this assumption should be true.

The modelling work here explores consequences of the above assumption and the uncertainty in electron dynamics on the gas phase side of the interface. To do this, a simple model 1D DC Argon discharge with a water anode is used. The purpose of the work is not to make definitive predictions about the behavior at the plasma-liquid interface but rather to present a range of results that may encompass the true physical behavior. Additionally, the authors hope that the research presented here may motivate deeper studies of the gas-phase side of the interface, whether it be through ab initio calculations or experimental techniques.

For this paper, both kinetic and thermodynamic descriptions of the electron density at the interface are considered. A description of the 1D fully-coupled plasma-liquid model is given in \cref{sect:model}. In \cref{sect:results} it is shown that by varying the interfacial electron surface loss coefficient in the kinetic description or a Henry's law like coefficient in the thermodynamic description, the electron density on the gas phase side of the interface can be changed by orders of magnitude. Moreover, if electrons coming from the bulk are not absorbed at the interface, they become thermalized through non-recombinatory collisions. Conclusions are given in \cref{sect:conclusions}. A brief description of the novel code used to implement the model is given in \cref{sect:appendix}. Anyone interested in downloading and using the code may access it at \url{https://github.com/lindsayad/zapdos}.

\section{Model Description}
\label{sect:model}

The fully coupled 1D plasma liquid model is implemented in a code developed by the authors. A brief description of the code, which is open source and free to use \cite{zapdosSite}, is given in \cref{sect:appendix}. In the model, a DC atmospheric pressure argon discharge impinges on a very thin water layer. The powered electrode is biased negatively, making it the cathode. From the plasma's perspective, the water surface is the anode. Only elastic collisions, ground state ionization, and ground state excitation are considered. The model governing equations are described below. Continuity equations based on the drift-diffusion approximation are solved for the electrons and ions:

\begin{equation}
  \frac{\partial n_i}{\partial t} + \nabla\cdot\vec{\Gamma_i} = S_{iz}
  \label{eq:ions}
\end{equation}
\begin{equation}
  \frac{\partial n_e}{\partial t} + \nabla\cdot\vec{\Gamma_e} = S_{iz}
  \label{eq:electrons}
\end{equation}
\begin{equation}
  \vec{\Gamma_i} = \mu_i\vec{E}n_i - D_i\nabla n_i
  \label{eq:gamma_i}
\end{equation}
\begin{equation}
  \vec{\Gamma_e} = \mu_e\vec{E}n_e - D_e\nabla n_e
  \label{eq:gamma_e}
\end{equation}
\begin{equation}
  S_{iz} = \alpha_{iz}\lvert\vec{\Gamma_e}\rvert
  \label{eq:source_term}
\end{equation}

where $\mu$ is the mobility, D the diffusivity, $\alpha_{iz}$ the Townsend ionization coefficient, $\Gamma$ the species flux, S$_{iz}$ the ionization source term, n the species density, and $\vec{E}$ the electric field, equal to $\nabla V$ where V is the potential. Poisson's equation is solved for the potential:

\begin{equation}
  -\nabla^2V = \frac{e\left(n_i - n_e\right)}{\epsilon_0}
  \label{eq:poisson}
\end{equation}

where e is the Coulombic charge and $\epsilon_0$ is the permittivity of free space. The equation for the electron energy is:

\begin{equation}
  \frac{\partial\left(n_e\epsilon\right)}{\partial t} + \nabla\cdot\vec{\Gamma_{\epsilon}} = -e\vec{\Gamma_e}\cdot\vec{E} - \lvert\vec{\Gamma_e}\rvert\left(\alpha_{iz}\epsilon_{iz} + \alpha_{ex}\epsilon_{ex} + 3\frac{m_e}{m_i}\alpha_{el}T_e\right)
  \label{eq:electron_energy}
\end{equation}
\begin{equation}
  \vec{\Gamma_{\epsilon}} = \frac{5}{3}\epsilon\vec{\Gamma_e} -\frac{5}{3}n_eD_e\nabla\epsilon
  \label{eq:gamme_eps}
\end{equation}

where $\epsilon$ is the mean electron energy, $\epsilon_{iz}$ the electron energy lost in an ionization collision, $\alpha_{ex}$ the Townsend excitation coefficient, $\epsilon_{ex}$ the electron energy lost in an excitation collision, m$_i$ and m$_e$ the ion and electron masses respectively, $\alpha_{el}$ the Townsend elastic collision coefficient, and T$_e$ the electron temperature, equal to $\frac{2}{3}\epsilon$.

Plasma boundary conditions at the cathode are based on the work in \cite{hagelaar2000boundary} and \cite{sakiyama2007nonthermal}. For ions, electrons, and the electron energy, the conditions are respectively:

\begin{equation}
  \vec{\Gamma_i}\cdot\vec{n} = \frac{1-r_i}{1+r_i}\left(\left(2a_i-1\right)\mu_i\vec{E}\cdot\vec{n}n_i + \frac{1}{2}v_{th,i}n_i\right)
  \label{eq:ion_bc}
\end{equation}
\begin{equation}
    \vec{\Gamma_e}\cdot\vec{n} = \frac{1-r_{dens}}{1+r_{dens}}\left(-\left(2a_e-1\right)\mu_e\vec{E}\cdot\vec{n}\left(n_e-n_{\gamma}\right) + \frac{1}{2}v_{th,e}\left(n_e-n_{\gamma}\right)\right) - \left(1-a_e\right)\gamma_p\vec{\Gamma_p}\cdot\vec{n}
  \label{eq:electron_bc}
\end{equation}
\begin{equation}
    \vec{\Gamma_{\epsilon}}\cdot\vec{n} = \frac{1-r_{en}}{1+r_{en}}\left(-\left(2a_e-1\right)\frac{5}{3}\mu_e\vec{E}\cdot\vec{n}\left(n_e\epsilon-n_{\gamma}\epsilon_{\gamma}\right) + \frac{5}{6}v_{th,e}\left(n_e\epsilon-n_{\gamma}\epsilon_{\gamma}\right)\right) - \frac{5}{3}\epsilon_{\gamma}\left(1-a_e\right)\gamma_p\vec{\Gamma_p}\cdot\vec{n}
  \label{eq:energy_bc}
\end{equation}

where $r_i$, $r_{dens}$, $r_{en}$ are the boundary reflection coefficients for ions, electrons, and electron energy respectively (more discussion on $r_{en}$ shortly), $\gamma_p$ is the secondary electron emission coefficient, $\epsilon_{\gamma}$ is the energy of the secondary electrons, $\vec{n}$ is the outward facing normal vector, and:

\begin{equation}
  a_k =
    \begin{cases}
      1, & sgn_k\mu_k\vec{E}\cdot\vec{n}>0 \\
      0, & sgn_k\mu_k\vec{E}\cdot\vec{n}\leq0
    \end{cases}
  \label{eq:a}
\end{equation}
\begin{equation}
  v_{th,k} = \sqrt{\frac{8T_k}{\pi m_k}}
  \label{eq:v_th}
\end{equation}
\begin{equation}
  n_{\gamma} = \left(1-a_e\right)\frac{\gamma_p\vec{\Gamma_p}\cdot\vec{n}}{\mu_e\vec{E}\cdot\vec{n}}
  \label{eq:n_gamma}
\end{equation}

where $v_{th,k}$ is the thermal velocity of species $k$ and $n_{\gamma}$ is the density of secondary electrons. All $r_k$'s are set to zero at the cathode. At the interface of the plasma with the liquid phase, the ion boundary condition is the same as for the cathode with $r_i=0$. For electrons in the gas phase two formulations are considered. The first is the kinetic formulation given by \cref{eq:electron_bc} where $r_{dens}$ is variable. The second is a thermodynamic formulation analogous to Henry's law where the ratio of the liquid phase electron density to the gas phase electron density is specified by a variable H (equivalent to a Henry's Law coefficient):

\begin{equation}
  Hn_{e,g} = n_{e,l}
  \label{eq:electron_bc_thermo}
\end{equation}

The electron energy interfacial condition is the kinetic one, see \cref{eq:energy_bc}. Though $r_{dens}$ (or $H$ for the thermodynamic electron BC) at the interface is varied in the results that follow, $r_{en}$ is held constant at 0for most simulations. This is done for the following physical reasoning. Electrons can either pass freely into the liquid phase, carrying their energy with them, or they can be reflected. If they are reflected, then it is reasonable to expect these electrons to lose their energy in surface collisions such as vibrational excitation of H$_2$O until they are incorporated into the liquid. Thus though some electrons coming from the bulk may be reflected, it may be reasonable to assume that all the electron energy coming from the bulk is absorbed by the interface. However, in the interest of covering all realms of possibility (perhaps most electron collisions at the interface are low-loss elastic collisions for example), a study is conducted in which the amount of energy absorbed/reflected by the interface is varied. This is done by changing $\gamma_{en}$. Note that in the plots and discussion to follow, the surface loss coefficients $\gamma_{dens}$ and $\gamma_{en}$ will often be used instead of the reflection coefficients $r_{dens}$ and $r_{en}$. The relationship between surface loss and reflection coefficients is simply $\gamma_k = 1 - r_k$.

The liquid phase electron density interfacial condition is given simply by the continuity of flux. At the bottom of the liquid, electrons are assumed to recombine or flow out at a rate equivalent to the advective flux.

For potential conditions, V is set to zero at the end of the liquid domain. At the cathode, Kirchoff's voltage law for a circuit including a ballast resistor yields:

\begin{equation}
  V_{source} + V_{cathode} = \left(e\vec{\Gamma_i} - e\vec{\Gamma_e}\right)AR
  \label{eq:cathode}
\end{equation}

where A is the cross-sectional area of the plasma and R is the ballast resistance.

\begin{figure}[htbp]
  \centering
  \includegraphics[width=.6\textwidth]{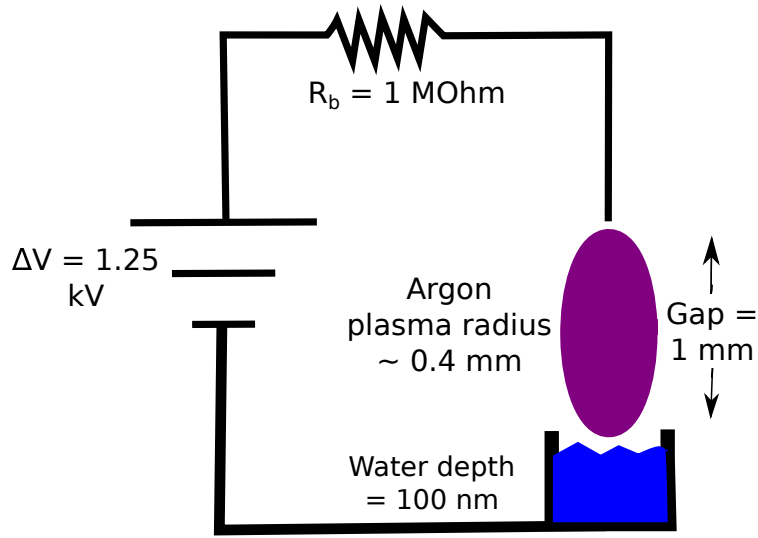}
  \caption{Circuit schematic of coupled plasma liquid system. Note that diagram is not to scale}
  \label{fig:circuit_PIC}
\end{figure}

Gas phase electron coefficients were calculated in the following way: Argon ionizization, excitation, and elastic collision cross sections were taken from the Phelps database \cite{yamabe1983measurement} at \cite{lxcat}. Then using the open source Boltzmann solver Bolos \cite{bolosSite} based on the work of Hagelaar \cite{hagelaar2005solving} electron energy distribution functions were calculated for 200 electric field points between $10^3$ and $10^7$ V/m. Then for each distribution function, $\mu_e$, D$_e$, $\epsilon$, and the necessary electron collision rate coefficients were calculated as defined by \cite{hagelaar2005solving}. Transport and rate cofficients were tabulated against the mean energy. These lookup-tables were then referenced during solution of the fluid equations. The details of the inputs for the fluid simulations are given in tables \ref{tab:params_PIC} and \ref{tab:coeffs_PIC} and figure \ref{fig:circuit_PIC}. Mesh sizes for the simulations were typically around 200 elements with most elements located in the cathode and interfacial regions. Each individual simulation took between 12 and 60 seconds to run.

\begin{table}[htpb]
  \begin{center}
    \begin{tabular}{c |c }
        Parameter & Value \\ \hline \hline
        Gas & Argon \\
        Pressure & 1 atm \\
        $\gamma_p$ & 0.15 \\
        A & $5.02\cdot10^{-7}m^{2}$ \\
        R & $10^6\Omega$ \\
        V$_{source}$ & 1.25 kV \\
        Gas Domain & 1 mm \\
        Liquid Domain & 100 nm \\
        $\epsilon_{\gamma}$ & 3 eV \\
        T$_i$ & 300 K \\
        \hline
    \end{tabular}
  \end{center}
  \caption{Plasma liquid simulation input parameters}
  \label{tab:params_PIC}
\end{table}

\begin{table}[htpb]
  \begin{center}
    \begin{tabular}{c |c |c }
      Coefficient & Value & Source\\ \hline \hline
      \rule{0pt}{3ex}$\mu_e$ & Variable & \cite{bolosSite} \\
      D$_e$ & Variable & \cite{bolosSite} \\
      $\mu_i$ & $3.52\cdot10^{-4} m^2s^{-1}V^{-1}$ & \cite{richards1987continuum} \\
      D$_i$ & $5.26\cdot10^{-6} m^2s^{-1}$ & \cite{richards1987continuum} \\
      $\alpha_{iz}$ & Variable & \cite{bolosSite} \\
      $\alpha_{ex}$ & Variable & \cite{bolosSite} \\
      $\alpha_{el}$ & Variable & \cite{bolosSite} \\
      $\epsilon_{iz}$ & 15.76 eV & \cite{lxcat} \\
      $\epsilon_{ex}$ & 11.5 eV & \cite{lxcat} \\
      \hline
    \end{tabular}
  \end{center}
  \caption{Plasma liquid simulation input parameters}
  \label{tab:coeffs_PIC}
\end{table}

\section{Results and Discussion}
\label{sect:results}

\begin{figure}[H]
  \centering
  \includegraphics[width=.75\textwidth]{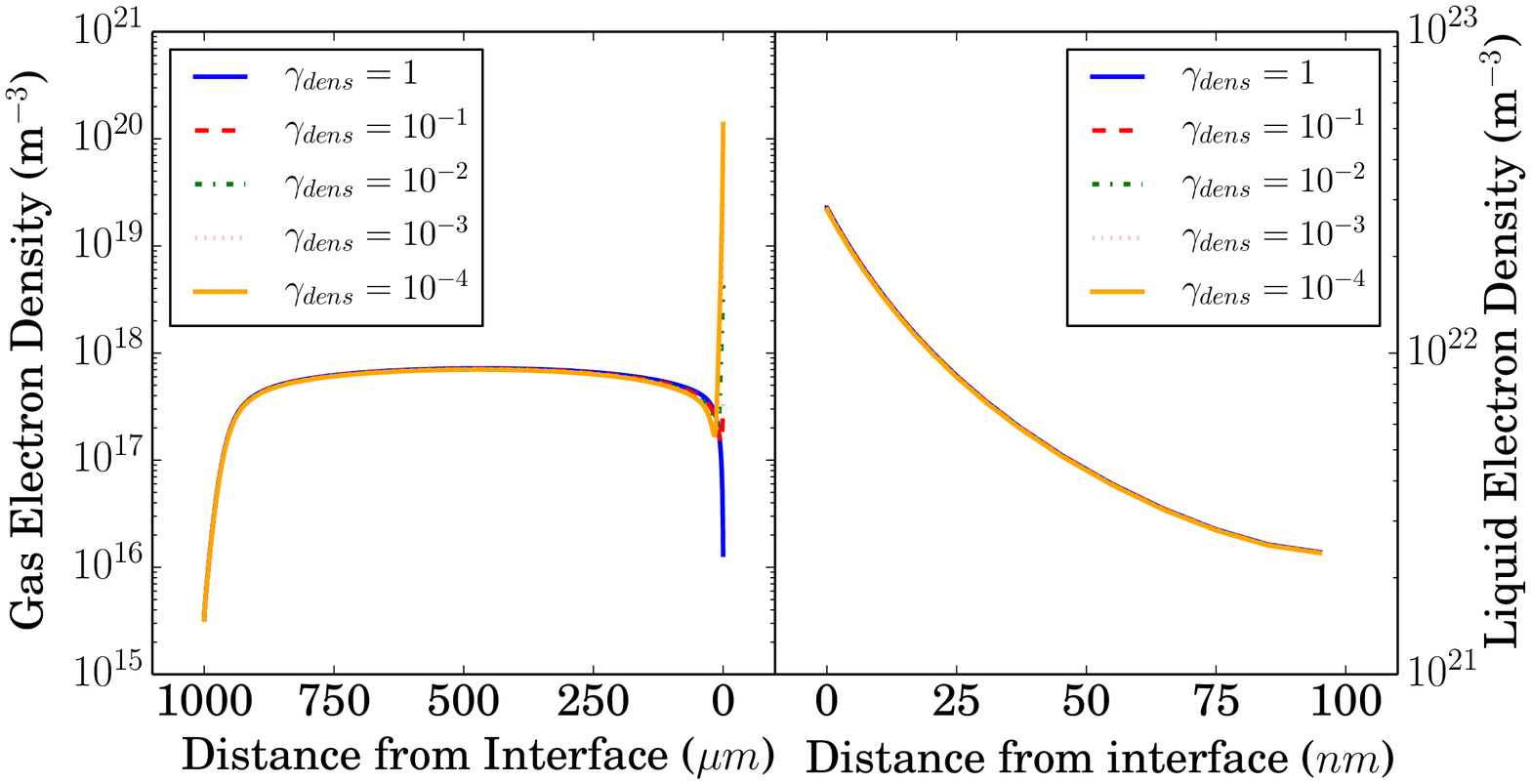}
  \caption{Electron density as a function of the interfacial surface loss coefficient}
  \label{fig:electrons}
\end{figure}

\begin{figure}[H]
  \centering
  \includegraphics[width=.75\textwidth]{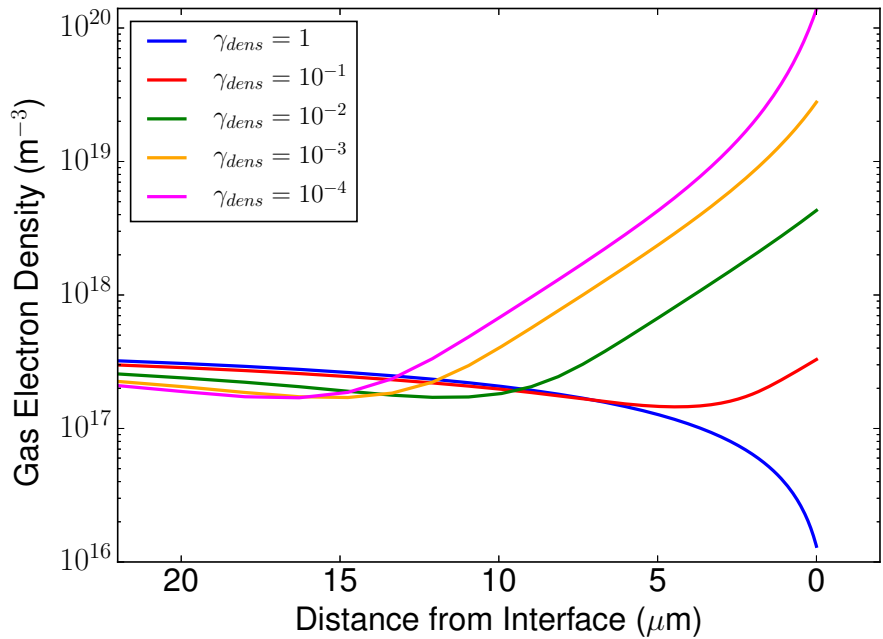}
  \caption{Electron density as a function of the interfacial surface loss coefficient. Final 20 $\mu$m of the gas phase before the interface.}
  \label{fig:electrons_int}
\end{figure}

\begin{figure}[H]
  \centering
  \includegraphics[width=.75\textwidth]{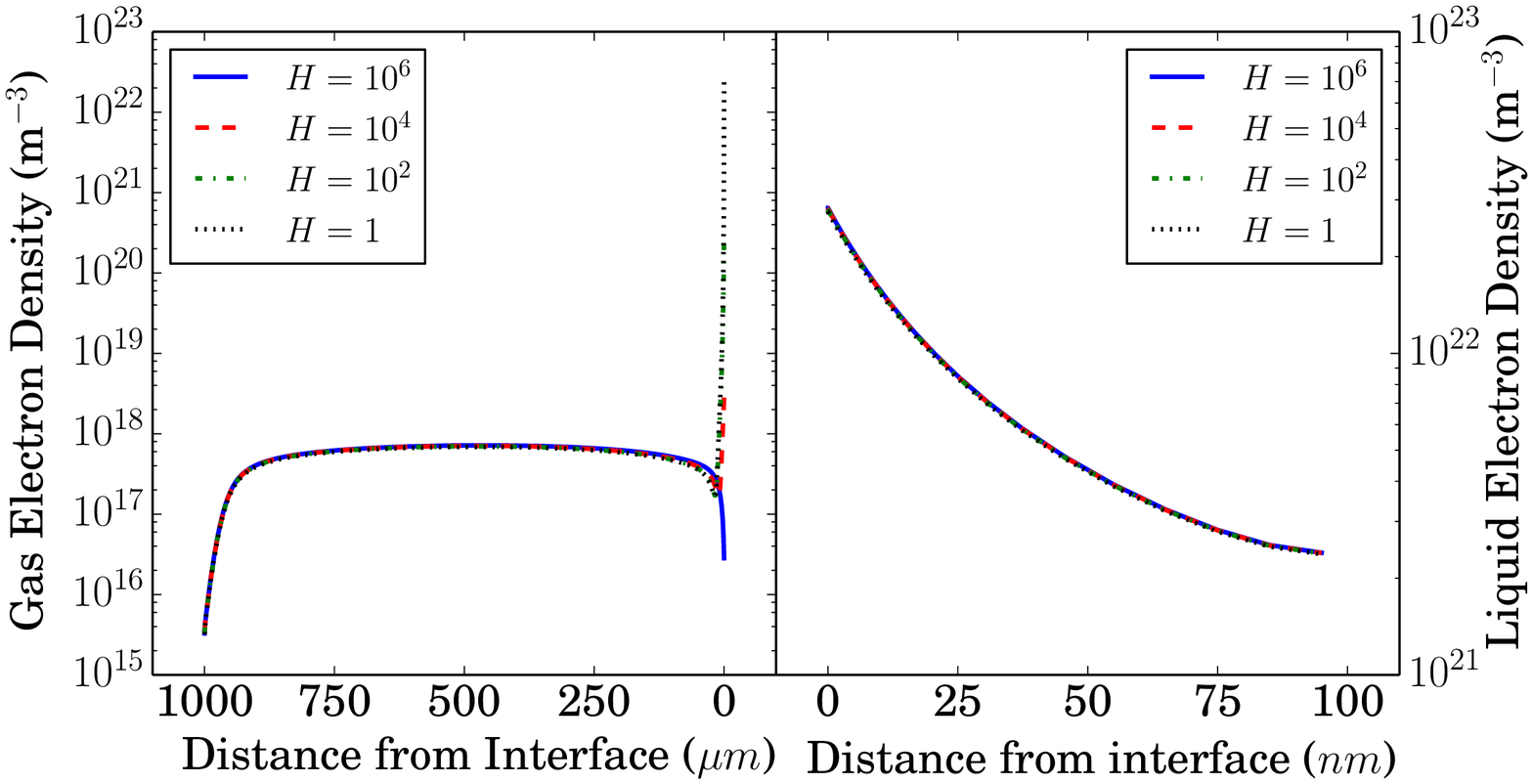}
  \caption{Electron density as a function of $H$ using the thermodynamic boundary condition. Shows same trend as \cref{fig:electrons}}
  \label{fig:electrons_thermo}
\end{figure}

\begin{figure}[H]
  \centering
  \includegraphics[width=.75\textwidth]{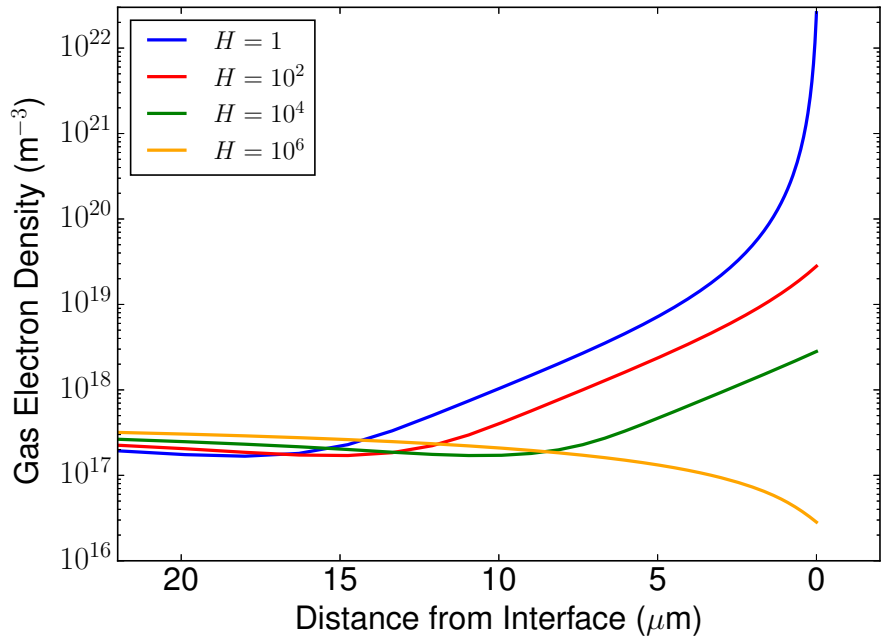}
  \caption{Electron density as a function of $H$ over the last 20 $\mu$m of the gas phase. Shows same trend as \cref{fig:electrons_int}}
  \label{fig:electrons_thermo_int}
\end{figure}

\Cref{fig:electrons} shows the electron density in both the gas and liquid phases as a function of the interfacial surface loss coefficient. The cathode and bulk profiles are unaffected by changing $\gamma_{dens}$. However, as one might expect, decreasing the surface loss coefficient leads to a build-up of electrons on the gas phase side of the interface, seen more clearly in \cref{fig:electrons_int}. Similar behavior can be achieved by decreasing the $H$ coefficient in \cref{eq:electron_bc_thermo,fig:electrons_thermo_int}. In order to observe anode characteristics akin to those for a plasma in contact with a metallic electrode ($\gamma_{dens}=1$), $H$ must be on the order of $10^6$. This is on the same order of magnitude as Henry's Law coefficients for H$_2$O$_2$ and HNO$_3$, both very hydrophilic species. If $H$ is reduced to $10^4$, the gas phase electron density near the interface increases by an order of magnitude. If $H$ is further reduced to $10^2$, only slightly less hydrophilic than OH, then the gas phase interfacial density rockets up to three orders of magnitude greater than the metallic anode base case. Decreasing $H$ further only continues the trend.

\begin{figure}[H]
  \centering
  \includegraphics[width=.75\textwidth]{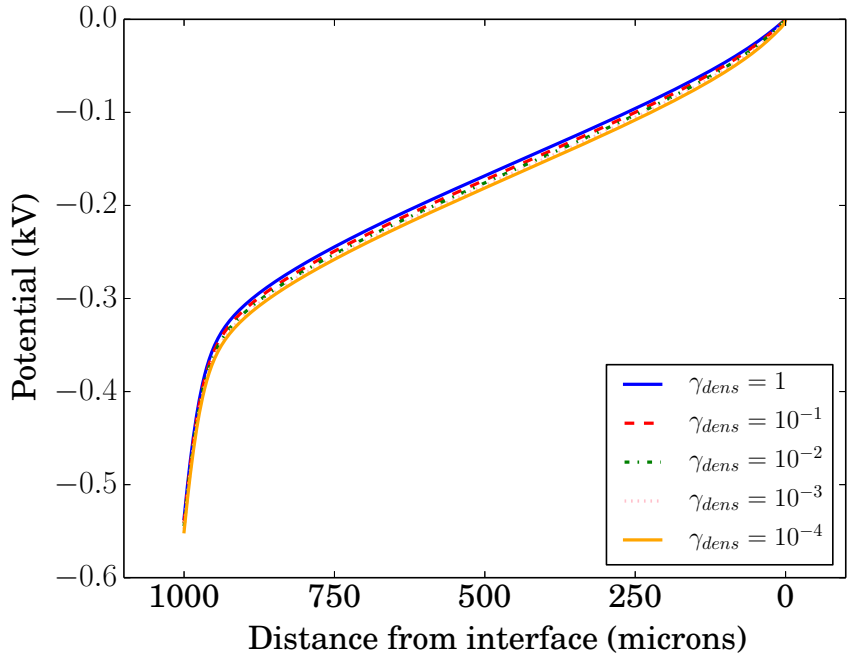}
  \caption{Potential as a function of the interfacial surface loss coefficient}
  \label{fig:potential}
\end{figure}

\begin{figure}[H]
  \centering
  \includegraphics[width=.75\textwidth]{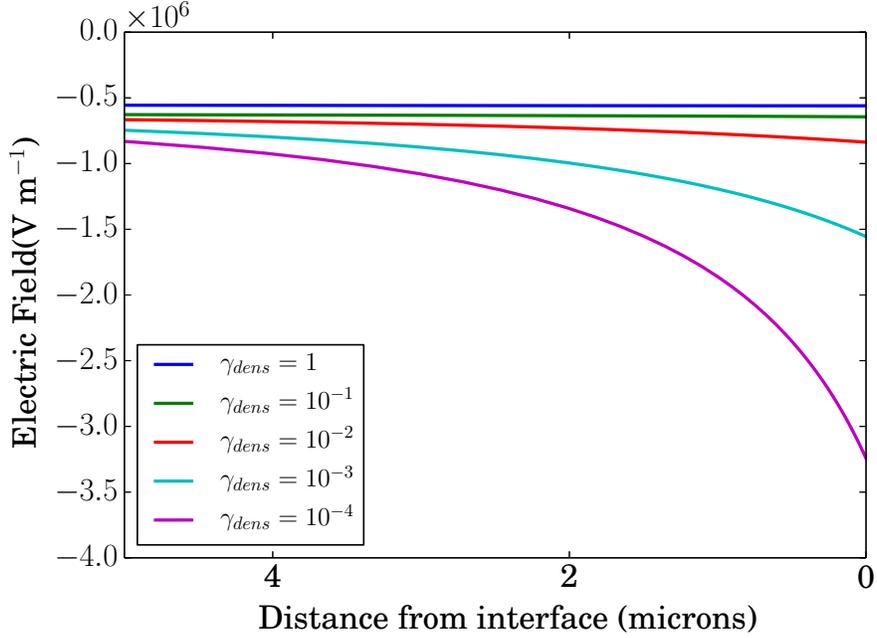}
  \caption{Electric field near the interface as a function of the interfacial surface loss coefficient}
  \label{fig:efield_int}
\end{figure}

Despite the dramatic functional dependence of the gas phase electron density in the anode, the liquid phase electron density profile remains unchanged as $\gamma_{dens}$ is varied. The reason for this can be seen by looking at \cref{fig:potential}. Like the liquid phase electron density profile, the potential drop across the plasma-liquid system is unaffected by changing $\gamma_{dens}$. This means that the system DC current is also unaffected, roughly 1000 Amps m$^{-2}$ for all simulation cases. Away from the cathode, all the current is carried by electrons, thus the electron current at the interface between the gas and liquid must also remain unchanged as $\gamma_{dens}$ is varied. With the liquid phase electron input thus unaffected by $\gamma_{dens}$, the liquid phase electron density profile remains constant. Varying $\gamma_{dens}$ does change the potential and electric field profiles near the interface; this is shown in \cref{fig:efield_int}. From the low reflection to high reflection extremes, the interfacial electric field increases by about a factor of seven.

\begin{figure}[H]
  \centering
  \includegraphics[width=.75\textwidth]{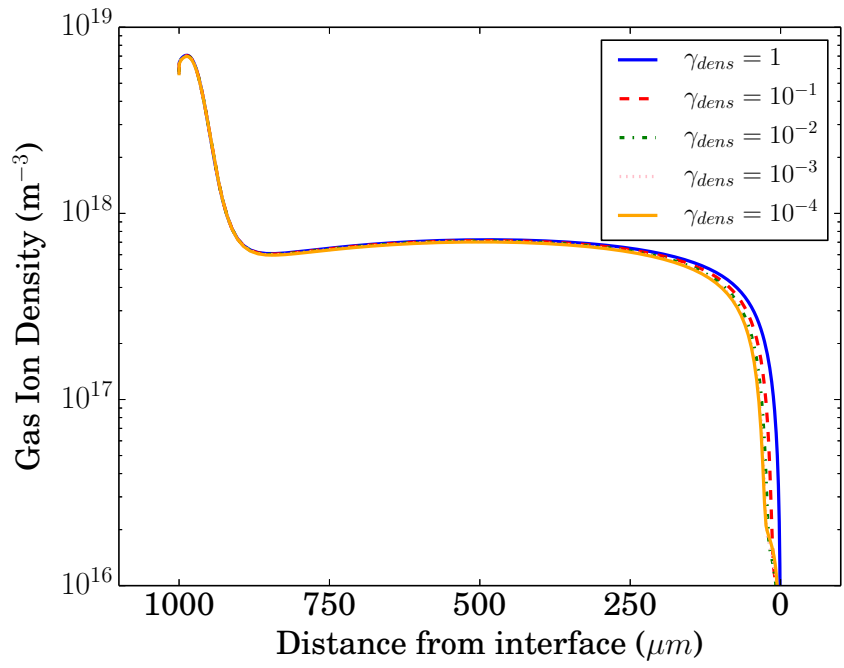}
  \caption{Ion density as a function of the interfacial surface loss coefficient}
  \label{fig:ions}
\end{figure}

\begin{figure}[H]
  \centering
  \includegraphics[width=.75\textwidth]{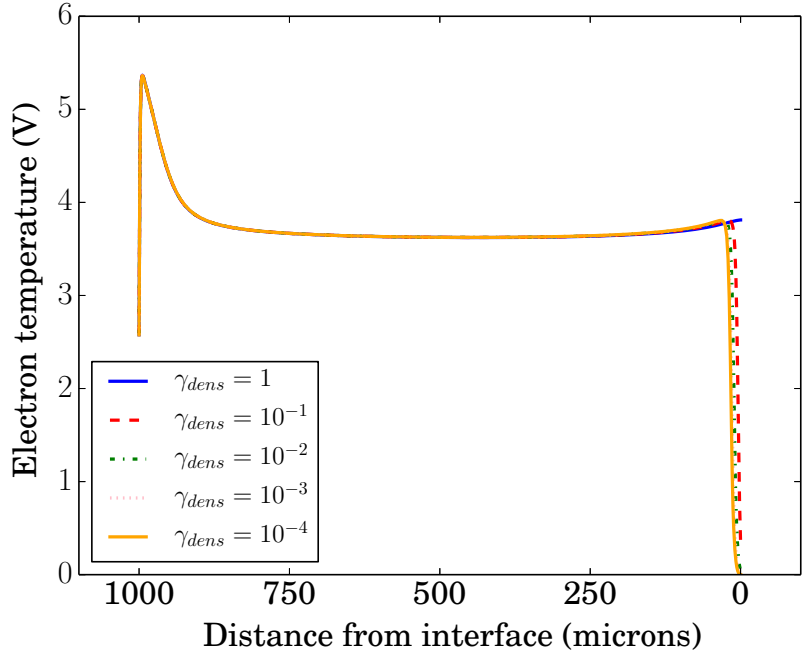}
  \caption{Electron temperature as a function of the interfacial surface loss coefficient}
  \label{fig:e_temp}
\end{figure}

As with the electron density, the cathode and bulk electron temperature profiles in \cref{fig:e_temp} do not change as $\gamma_{dens}$ is varied. However, there is major variation in the anode. This variation arises from the assumption described in the model description section that electrons coming from the bulk either carry their energy into the liquid phase upon absorption or else if reflected lose their energy through interfacial surface collisions. The greater the reflection, the lower the average energy of electrons near the interface because of non-recombinatory surface collisions. This is what is observed in \cref{fig:e_temp}. This trend in electron energy also explains the slight variation in anode ion density profiles seen in \cref{fig:ions}. Lower electron mean energy near the interface means a smaller fraction of electrons with sufficient energy to create ionization and a smaller Townsend ionization coefficient. Because in this model ionization is proportional to the electron flux magnitude and because the electron flux magnitude is constant with respecto to $\gamma_{dens}$, the decrease in $\alpha_{iz}$ corresponds to a decrease in the rate of ionization. Hence the ion density rises to its bulk value farther from the anode for decreasing $\gamma_{dens}$.

\begin{figure}[H]
  \centering
  \includegraphics[width=.75\textwidth]{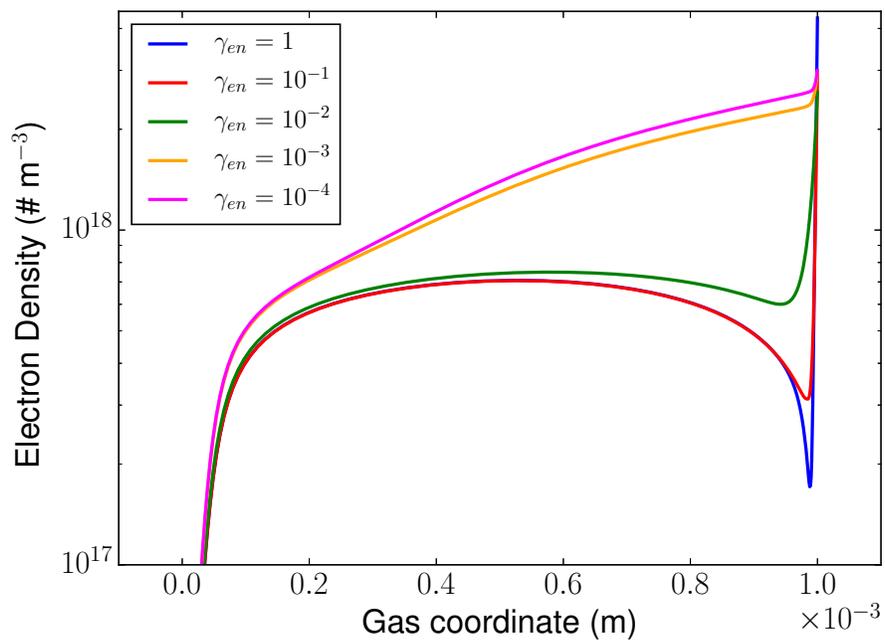}
  \caption{Gas phase electron density as a function of the electron energy interfacial surface loss coefficient. ($\gamma_{dens} = 10^{-2}$ for all cases)}
  \label{fig:electrons_en_sweep}
\end{figure}

\begin{figure}[H]
  \centering
  \includegraphics[width=.75\textwidth]{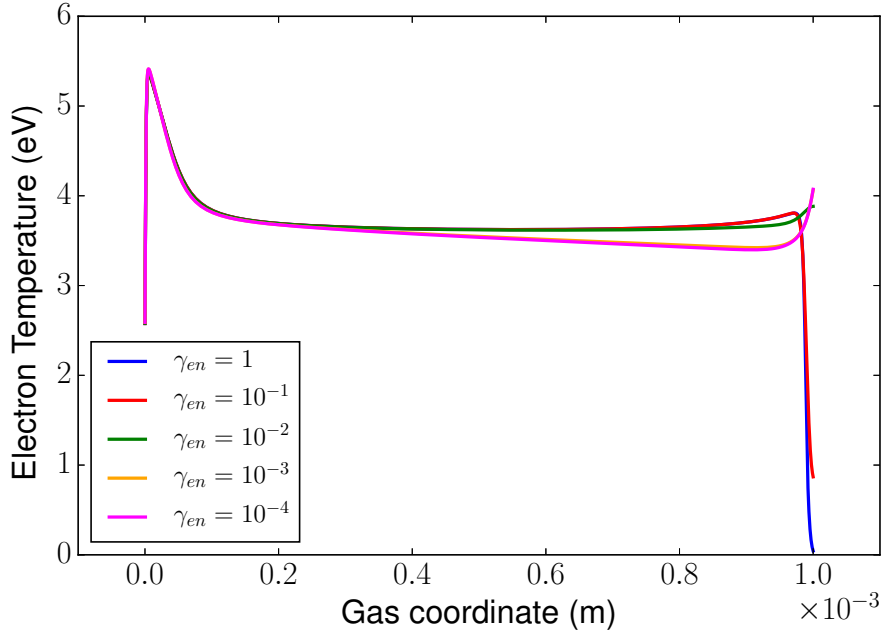}
  \caption{Gas phase electron temperature as a function of the electron energy interfacial surface loss coefficient. ($\gamma_{dens} = 10^{-2}$ for all cases)}
  \label{fig:etemp_en_sweep}
\end{figure}

The physically correct boundary condition for the electron energy at the interface is unknown. However, we can vary the amount of electron energy that is absorbed/reflected at the interface and see whether that affects the most important result of the above figures: that interfacial electron density increases significantly as the electron surface loss coefficient is decreased. \Cref{fig:electrons_en_sweep} shows the effect of varying the amount of energy lost at the interface when $\gamma_{dens}$ is kept constant at $10^{-2}$. A couple of trends are notable. The first is that as the energy reflection is increased, e.g. as $\gamma_{en}$ is decreased, the bulk electron density increases; moreover, instead of retaining a flat profile through the bulk, the electron density increases almost linearly moving from cathode to anode. Additionally, as $\gamma_{en}$ decreases the jump in electron density at the anode/interface decreases. The combination of these effects results in anodic electron densities that differ by less than a factor of two over values of $\gamma_{en}$ that span four orders of magnitude. Moreover, no matter the value of $\gamma_{en}$, the anodic electron density with $\gamma_{dens} = 10^{-2}$ is over an order of magnitude higher than if the surface loss coefficients for electrons is set to unity. Thus, we conclude that the important result of increasing anodic electron density with decreasing $\gamma_{dens}$ is relatively insensitive to the choice of $\gamma_{en}$; e.g. without knowing how to properly handle the electron energy boundary condition at the interface, we can still reasonably conclude that a decreasing surface loss coefficient will significantly increase the density of gas phase electron at the interface. The effect of varying $\gamma_{en}$ on the electron temperature gas phase profile is shown in \cref{fig:etemp_en_sweep}. Changes in the cathode and bulk profiles are minimal. However, as one might intuitively expect, increasing energy reflection increases the anodic electron temperature. An increase in electron temperature from the bulk to the anode (observed for $\gamma_{en} = 10^{-4}$) is more consistent with high current atmospheric argon PIC simulations. \cite{emiComm}

These trends in the anode electron density and electron temperature at the anode could play an important role in more complex models that consider evaporation of H$_2$O and dilute aqueous species. The rates of reactions of electrons with these species will depend strongly on the electron density and the electron energy distribution. Different energy distributions might favor vibrational excitation of H$_2$O or dissociative attachment and the production of electronegative plasma species like O$^-$ and OH$^-$. The near interface gas chemistry will of course couple back into the liquid phase chemistry. Future work with more complex models will investigate how changing $\gamma_{dens}$ and $\gamma_{en}$ affects plasma and liquid chemistry. However, in order to limit the scope of possible results and increase the predictive capability of such models, there must be more certainty in interfacial parameters like $\gamma_{dens}$ and in the interfacial energy dynamics (represented in this work by $\gamma_{en}$. Determination of such characteristics will likely require finer scale simulations (molecular dynamics for instance) and/or new experimental diagnostics that are capable of probing near-interface gas dynamics.

\section{Conclusions and Future Work}
\label{sect:conclusions}

In this work it is found that varying the electron surface loss coefficient at the plasma-liquid interface can have significant impacts on both the electron density and electron energy near-interface characteristics. Future work will investigate how these variations could impact plasma chemistry arising from the interaction of the near-interface gas electrons with volatile chemical species coming from the liquid phase. Additionally the model will be expanded to multiple dimensions in the hopes of reproducing the spreading of discharges over the liquid surface as a function of solution conductivity. \cite{rumbach2015solvation} Finally, finer scale molecular simulations and/or experiments must be conducted in order to understand the true physical behavior of electrons in the gas near the interface and to accurately determine fluid modelling parameters like $\gamma_{dens}$.

\section{Acknowledgments}

The authors acknowledge support from the United States Department of Energy, Office of Fusion Science Plasma Science Center and from the National Science Foundation. Thanks to Emi Kawamura for providing PIC simulation results that helped us in validation of the fluid model implementation. Thanks to Ranga Gopalakrishnan for his literature search of aqueous reactions, rate constants, and transport coefficients. Finally, a very large thanks to the Moose team for all the help along the way in building Zapdos into a successful code.

\FloatBarrier

\bibliographystyle{unsrt}
\bibliography{plasliq_paper}

\section{Appendix: Zapdos Code Description}
\label{sect:appendix}

Zapdos is built on top of the Multiphysics Object-Oriented Simulation Environment (MOOSE) \cite{mooseSite} and libMesh \cite{libmeshSite} codes. MOOSE employs finite element methods (Continuous Galerkin, Discontinuous Galerkin, or a combination) to solve fully coupled (or segregated through the use of MultiApps) systems of partial differential equations (PDEs). After using FEM to discretize the governing equations, MOOSE interfaces with the code PetSc \cite{petscSite} to solve the (non-)linear system of algebraic equations via Newton's method globalized through a line search:

\begin{equation}
  \tilde{J}(\vec{u}^k)\vec{\delta u}^k = -\vec{R}(\vec{u}^k)
  \label{eq:Newton}
\end{equation}
\begin{equation}
  \vec{u}^{k+1} = \vec{u}^k + s\vec{\delta u}^k
  \label{eq:line_search}
\end{equation}

where $\vec{u}^k$ is the solution vector for iterate $k$, $\vec{R}$ is the residual vector, and $\tilde{J}$ is the Jacobian matrix formed by taking the derivatives of the residual vector with respect to the solution vector.\cite{knoll2004jacobian} \Cref{eq:Newton} may be solved through either direct or iterative methods (usually GMRES with a variety of preconditioning methods including incomplte lower-upper, block jacobi, additive Schwartz, (geometric) algebraic multigrid, etc.). Line search techniques (\cref{eq:line_search}) are based on the methods in \cite{dennis1996numerical}. For application programmers building on top of MOOSE, it is their responsibility to code the residual and Jacobian statements that represent their physics. Residual statements are pieces of the physical governing equations cast in the weak form. A maximally efficient application code in terms of computational time will have a complete and correct set of Jacobian statements corresponding to derivatives of the residuals with respect to the solution variables and will employ the standard Newton method plus line search. If developer time is at a premium, some Jacobian statements can be incomplete or omitted and a Jacobian-Free Newton-Krylov (JFNK) method can be employed in the stead of standard Newton. However, this comes at the cost of computational effiency. The low-temperature plasma application Zapdos has been designed with the former strategy in mind: complete and correct Jacobian statements so that standard Newton can be used. As Zapdos is developed, new pieces of physics with new analytical Jacobians are compared against PetSc Jacobians formed through finite differencing of the residual statements to ensure accuracy.

Zapdos partitions governing equation terms into individual pieces called kernels. Each kernel contains the residual (simply the term cast in weak form) and the corresponding Jacobian statements. Consider the drift flux term in charged particle continuity equations: $\nabla\cdot\left(-sgn(q)\mu\nabla V\right)$. After casting into the weak form and taking the volume term, the corresponding Zapdos code looks like:

\begin{lstlisting}[language=C++]
Real EFieldAdvection::computeQpResidual()
{
  return _mu[_qp] * _sign[_qp] * std::exp(_u[_qp]) * -_grad_potential[_qp] * -_grad_test[_i][_qp];
}

Real EFieldAdvection::computeQpJacobian()
{
  return _mu[_qp] * _sign[_qp] * std::exp(_u[_qp]) * _phi[_j][_qp] * -_grad_potential[_qp] * -_grad_test[_i][_qp];
}

Real EFieldAdvection::computeQpOffDiagJacobian(unsigned int jvar)
{
  if (jvar == _potential_id)
    return _mu[_qp] * _sign[_qp] * std::exp(_u[_qp]) * -_grad_phi[_j][_qp] * -_grad_test[_i][_qp];
  else
    return 0.;
}
\end{lstlisting}

where \_u is the solution variable that the kernel is applied to (could be any ion species or electron), \_phi and \_test represent finite element shape functions (\_phi = \_test in all cases if using the same order and family of shape functions for all solution variables), and \_qp represent the positions of quadrature points. By splitting governing equations in this way into individual terms/kernels, code reproduction is kept at a minimum; analagous terms can be used in many different settings, e.g. a ``diffusion'' term has the exact same mathematical form as a ``conduction'' or ``viscosity'' term and so the same kernel code can be used for all three physics cases. Material properties like mobilty and diffusivity are defined in a materials file separated from the kernel code. Material properties can be defined as constants, as functions of the solution variables, or as properties to be read from look-up tables. Through MOOSE, Zapdos provides an interface for linear, bilinear, and spline interpolation of material properties. Boundary conditions are available in ``Nodal'' and ``Integrated'' flavors. Nodal boundary conditions are dirichlet like conditions that are enforced strongly. Integrated boundary conditions are cast in the weak form and often arise from performing integration by parts on divergence terms in the governing equations.

At the time of writing Zapdos has the necessary kernels and boundary conditions for solving gas phase DC discharge fluid models as well as conventional convection-diffusion-reaction equations for dilute species in a fluid (a future publication will demonstrate fully-coupled simulation of a DC discharge impinging on a liquid surface). Another student is working on implementing RF plasma simulation capabilities (for capacitively coupled plasmas this will only require slight modification of some boundary conditions; inductively coupled plasmas will require a little more work).

Zapdos solutions are output to an exodus file by default, although MOOSE provides varying levels of support for some other output file formats (including full support for simple CSV). These exodus files are then most commonly viewed graphically with either of the free and open source packages Visit or Paraview. For users more programatically inclined, Paraview provides python tools that enable the user to directly read the exodus file and create publication level plots in MatPlotLib with a single script (as is done for most of the figures in this paper). For transient simulations, results for any solution or auxiliary variable can be viewed while the calculation is on-line. Results are also not lost if a solve is cancelled for any reason. These features enable quick convergence debugging of a failing or failed solve.

\begin{figure}[htbp]
  \centering
  \includegraphics[width=.4\textwidth]{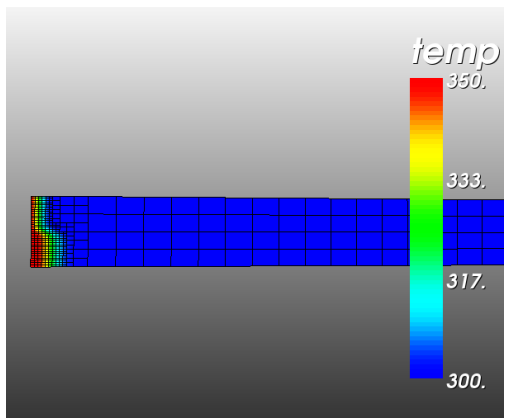}
  \caption{Propagating front. Time step 15. Note how the mesh is fine around the solution gradients and coarse elsewhere.}
  \label{fig:step15}
\end{figure}

\begin{figure}[htbp]
  \centering
  \includegraphics[width=.4\textwidth]{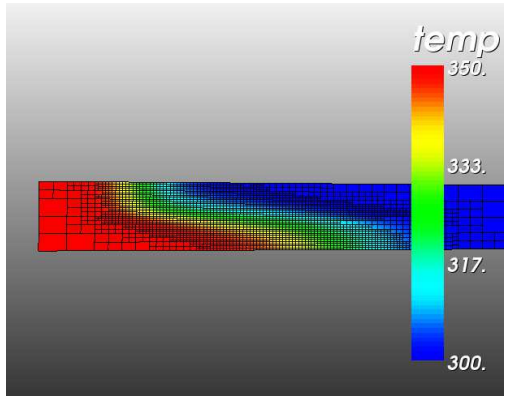}
  \caption{Propagating front. Time step 49. Note how the mesh is fine around the solution gradients and coarse elsewhere.}
  \label{fig:step49}
\end{figure}

The final feature of Zapdos worth mentioning is the adaptive mesh refinement inherited from MOOSE. The user can choose from several different indicators, including the jump in a solution gradient or laplacian between elements, for determing where mesh refinement should take place. Figures \ref{fig:step15} and \ref{fig:step49} show the propagation of a front through a domain in which the top and bottom halves have different mobilities. The mesh tracks with the head of the front; the mesh is finer in regions of steeper gradients. This feature can be incredibly useful when trying to track ionization bullets or similar phenomena.

\end{document}